\documentclass{article}
\usepackage{graphicx}
\usepackage{amsmath}
\usepackage{amsfonts}
\usepackage{amssymb}

\begin{document}
\bigskip

\begin{center}
{\large \textbf{Alternative Hamiltonian Descriptions and Statistical
Mechanics}}

\vskip1cm

%
%
%
%
%
%
%
%
\noindent{E. Ercolessi$^{1}$, G. Marmo$^{2}$ and G. Morandi$^{1}$ \vskip0.5cm
\noindent\textit{{$^{1}$ Dipartimento di Fisica, Universita' di Bologna, INFM
and INFN,\\[0pt]Viale Berti-Pichat 6/2, I-40127 Bologna, Italy\\[0pt]$^{2}$
Dipartimento di Scienze Fisiche, Universita' di Napoli Federico II and INFN,
\\[0pt]Via Cinzia, I-80126 Napoli, Italy\\[0pt]}} }

\vskip2cm
%
%
\today

\abstract
{We argue here that, just as it happens in Classical and Quantum Mechanics,
where it has been proven that alternative Hamiltonian descriptions can be
compatible with a given set of equations of motion, the same holds true in
the realm of Statistical Mechanics, i.e. that alternative Hamiltonian
descriptions do lead to the same thermodynamical description of any
physical system. } \newpage
\end{center}

In recent years, the research activity in the field of Theoretical Mechanics
has brought to the fore the previously unexpected fact that large classes of
(classical) dynamical systems can admit of genuinely inequivalent Lagrangian
$\cite{MFLMR}$ and Hamiltonian $\cite{AH}$ descriptions. By ''genuinely'' we
mean here, e.g., Lagrangians that do not differ merely by a so-called total
time derivative. As a standard example of this sort we may quote, e.g., the
$n$-dimensional isotropic harmonic oscillator with, for simplicity, unit mass
and frequency for which any Hamiltonian of the form:
\begin{equation}
\mathcal{H}_{B}=\frac{1}{2}B_{ij}(\delta^{ik}p_{k}\delta^{jl}p_{l}+q^{i}q^{j})
\end{equation}
with $B$ a symmetric and nondegenerate $n{\times} n$ real matrix with constant
entries is an admissible Hamiltonian. Similar examples can be exhibited within
the Lagrangian formulation of Classical Dynamics as well as at the quantum
level $\cite{DMS90}$. To the best of our knowledge the last case was first
considered by E.P.Wigner$\cite{W50}$ back in 1950.

The same dynamical evolution can then be compatible with quite different
Lagrangian and Hamiltonian formulations and the latter, useful and powerful as
they may be, appear to some extent as ''superstructures'' that can be used
interchangeably to describe one and the same dynamical system.

The situation seems to change rather drastically when one moves to Statistical
Mechanics, however. Indeed, in their standard formulation $\cite{TO}$, both
Classical and Quantum Statistical Mechanics appear to be deeply rooted in the
Hamiltonian formalism, and the basic quantity out of which one can deduce all
the thermodynamic functions that, at equilibrium at least, are the basic
observable and measurable quantities, namely the partition function, appears
to require the specification of a Hamiltonian and of a symplectic volume in
phase space at the classical level or of a Hamiltonian operator and of an
Hermitian scalar product in the Hilbert space at the quantum one.

We will argue in this Note that this not quite the case and that (with some
necessary technical limitations) alternative Hamiltonian descriptions all
yield the same partition function, and hence the same expression for all the
observable quantities that can be associated with a given dynamical system.

The general framework we will begin considering is that of the symplectic
formulation of Classical Dynamics $\cite{MSSZ}$.We start then with a
configuration space $\mathbb{Q}$, assumed to be a finite-dimensional smooth
manifold, and with a symplectic form $\Omega$ on the cotangent bundle
$\mathbb{T}^{\ast}\mathbb{Q}.$ Given any (smooth) function $H\in
\mathcal{F}(\mathbb{Q)}$, the pair $(\Omega,H)$ will determine uniquely the
vector field $\Gamma\in\mathcal{X}(\mathbb{T}^{\ast}\mathbb{Q}$ $)$ such that:%

\begin{equation}
i_{\Gamma}\Omega=-dH
\end{equation}
\ Alternatively, the pair $(\Omega,\Gamma)$ will determine the Hamiltonian $H$
upto an additive constant.

We will stick mostly in what follows to the one-dimensional case and to the
case in which $H$ is bounded from below (and we will assume then, with no loss
of generality: $H\geq0$) and the energy ''surfaces'' are compact. Hence the
orbits are closed lines in the two-dimensional phase space. Under these
assumptions, the dynamical system associated with the pair $(H,\Omega=dp\wedge
dq)$ will be Liouville-integrable $\cite{Ar}$ and will be described by a
single pair $(J,\phi)$ of action-angle variables. In terms of the latter:
$H=H(J),d\phi/dt=\nu\doteq dH/dJ,\Omega=dJ\wedge d\phi$ and the dynamical
vector field $\Gamma$ will be expressed as: $\Gamma=\nu\partial/\partial\phi$.

It is immediate to see that, in the manifold obtained by excluding from the
configuration space the critical points of $H$, the symplectic form $\Omega$
can also be written as:%
\begin{equation}
\Omega=dH\wedge\xi
\end{equation}
where:%
\begin{equation}
\xi=d(\frac{\phi}{\nu})
\end{equation}
is \ a closed one-form such that: $i_{\Gamma}\xi=1$. Locally at least, then,
$\xi=dt$ and this will define for us\footnote{Note that $\xi$ is not unique.
Indeed $\xi$ and $\xi+\alpha dH$, with $\alpha=\alpha(H)$ will be just as
good. The new one-form will be again closed and it will lead simply to a shift
in the origin of the ''time function''.} a ''time function'' $t$.

%
%
%
%
Consider as an example a one-dimensional harmonic oscillator with mass \ $m$
and proper frequency $\omega$, with the standard Hamiltonian:
\begin{equation}
H=\frac{1}{2}{\Bigg\{}\frac{p^{2}}{m}+m\omega^{2}q^{2}{\Bigg\}}%
\end{equation}
Then, on $\mathbb{R}^{2}-\{\mathbf{0}\}$ (the origin being the only critical
point of $H$) and, with the standard symplectic form $\Omega=dp\wedge dq$, the
vector field will be:
\begin{equation}
\Gamma=\frac{p}{m}\frac{\partial}{\partial q}-m\omega^{2}q\frac{\partial
}{\partial p}%
\end{equation}
%
%
%
%

Then:
\begin{equation}
\Omega=dp\wedge dq=dH\wedge\xi
\end{equation}
with (locally, of course):
\begin{equation}
\xi=dt=\frac{pdq-qdp}{2H}%
\end{equation}
(again ''modulo'' a multiple of $dH$), and the ''time function'' $t$ will be
given by: $t=(1/\omega)\tan^{-1}\{m\omega q/p\}$, which emphasizes its local character.

$\mathbb{R}^{2}-\{\mathbf{0}\}$ can be identified with $\mathbb{S}^{1}{{\times}
}\mathbb{R}^{+}$ parametrized by $dH$ and $dt$. Notice that, denoting by
$\Sigma(E)$ the one-dimensional ''surface'' of constant energy $E$:
\begin{equation}
\int\limits_{\Sigma(E)}dt=\frac{2\pi}{\omega}\doteq\tau
\end{equation}
with $\tau$ the period which, in this specific case, turns out to be a constant independent of the energy. In
more general cases $\tau=\tau(E)$ will be a function of the energy.The associated canonical\footnote{We will
restrict here to the canonical ensemble of (both classical and quantum) Statistical Mechanics.} partition
function is easily evaluated, and the well-known result$\cite{MO95}$ is:
\begin{equation}
\mathcal{Z}=h^{-1}\int\limits_{\mathbb{R}^{2}}\exp\{-\beta H\}\Omega
=h^{-1}\int\limits_{0}^{\infty}dE\exp\{-\beta E)\int\limits_{\Sigma
(E)}dt=\frac{1}{\beta\hbar\omega}%
\end{equation}
Here $\beta=1/k_{B}T$ with $T$ the (absolute) temperature and $k_{B}$ the
Boltzmann constant, while $h$ (and: $\hbar=h/2\pi$) is a numerically
undetermined constant with the dimension of an action\footnote{That can be
interpreted as the volume of some elementary cell in phase space.} that one is
forced $\cite{MO95}$ to introduce in the context of classical Statistical
Mechanics in order to obtain a dimensionless expression for the partition
function, so as to make sense of expressions such as : $\mathcal{F}%
=-\beta^{-1}\ln\mathcal{Z}$ for the (Helmoltz) free energy. As is also well
known, the value of $h$ is fixed unambiguously at that of Planck's constant at
the quantum level of Statistical Mechanics.

Alternative Hamiltonian descriptions are obtained $\cite{DMS90}$, for a
fixed vector field $\Gamma$, i.e. for a given set of equations of motion,
by changing together $H$ and the symplectic form $\Omega$ on
$\mathbb{T}^{\ast}\mathbb{Q}$ ($\mathbb{Q=R}$ in our case) in such a way
that $\Gamma$ be again Hamiltonian w.r.t. the new Hamiltonian and
symplectic form. Being one-dimensional (and Hamiltonian, hence
conservative) implies for our system that any alternative Hamiltonians must
necessarily be functions of the initial one (and of each other, of course).

In order to keep track of the correct dimensions of the various physical
quantities involved, let's consider a new Hamiltonian of the form:
\begin{equation}
H_{\phi}=\phi(H)\doteq\beta_{0}^{-1}f(\beta_{0}H)
\end{equation}
where $\beta_{0}$ is a ''fiducial'' quantity, fixed once and for all and having dimension $[energy]^{-1}$, and
$f(.)$ is a real function. We will assume $f^{\prime}>0$ throughout, and that in order: $i)$ to give a sensible
meaning to integrals (see below) over phase space and: $ii)$ not to change the number of critical points. The
original Hamiltonian will correspond of course to $f(x)=x$.

%
%
%
%
It is easy to prove that if $\Gamma$ is Hamiltonian w.r.t. $(H,\Omega)$, then
it will be Hamiltonian as well w.r.t. $(H_{\phi},\Omega_{\phi})$ (i.e.:
$i_{\Gamma}\Omega_{\phi}=-dH_{\phi}$), where $\Omega_{\phi}$ is defined as:
\begin{equation}
\Omega_{\phi}=dH_{\phi}\wedge dt
\end{equation}
%
%
%
%

Having redefined (through the new symplectic form) the volume element in phase
space, it is natural to redefine the partition function as:
\begin{equation}
\mathcal{Z}_{\phi}=h^{-1}\int\limits_{\mathbb{R}^{2}}\exp\{-\beta H_{\phi
}\}\Omega_{\phi}%
\end{equation}
But then:
\begin{equation}
\mathcal{Z}_{\phi}=-(h\beta)^{-1}\int d\exp\{-\beta E_{\phi}\}\int
\limits_{\Sigma(E_{\phi})}dt
\end{equation}

For the harmonic oscillator it is immediate to see that, again: $\mathcal{Z}%
_{\phi}=1/\beta\hbar\omega$, i.e. that the partition function is insensible to
the form of the Hamiltonian \textit{as long as the equations of motion
remain the same}.

As another example, consider the following nonlinear (and noncanonical) change
of coordinates for the linear oscillator $\cite{MSV}$(we set here for
simplicity: $m=\omega=1$):
\begin{equation}
Q=q(1+\phi),\text{ \ }P=p(1+\phi)
\end{equation}
where: $\phi=\phi(H)=f(\beta_{0}H)$ \ (cfr. Eq.$(11)$) and, in our units:
$H=(p^{2}+q^{2})/2$. A straightforward calculation shows that $\Gamma
=p\partial/\partial q-q\partial/\partial dp$, which is Hamiltonian w.r.t. the
standard pair $(H,\Omega=dp\wedge dq)$, is also Hamiltonian w.r.t. the pair
$(H^{\prime},\Omega^{\prime})$, where:
\begin{equation}
\Omega^{\prime}=dP\wedge dQ,\text{ \ }H^{\prime}=(1+\phi)^{2}H\equiv\frac
{1}{2}(P^{2}+Q^{2})
\end{equation}
Explicitly:
\begin{equation}
\Omega^{\prime}=F(H)dp\wedge dq
\end{equation}
with:
\begin{equation}
F(H)=(1+\phi)\{1+\phi+2H\phi^{\prime}\}\equiv\frac{dH^{\prime}}{dH}%
\end{equation}

By writing the symplectic volumes as: $\Omega=dH\wedge dt$ and: $\Omega
^{\prime} =dH^{\prime}\wedge dt$ respectively\footnote{This is possible with
the \textbf{same} time one-form $dt$ precisely because the dynamics is
unchanged.} and recasting Eq.$(23)$ in the form:
\begin{equation}
\mathcal{Z}=-\frac{1}{\beta h}\int d\exp(-\beta E)\int\limits_{\Sigma(E)}dt
\end{equation}
it is immediate to see that:
\begin{equation}
\int\exp(-\beta H^{\prime})dP\wedge dQ=\int\exp(-\beta H)dp\wedge dq
\end{equation}
i.e. that the partition function remains the same.

\bigskip

These results can be easily generalized. For a general
Hamiltonian\footnote{Again in one degree of freedom.} $H=H(q,p)$ with compact
energy ''surfaces'' and again endowed with a ''time one-form'' $dt$ (which we
know is granted) we can define again a period ( energy-dependent in general)
$\tau(E)$ as:
\begin{equation}
\tau(E)=\int\limits_{\Sigma(E)}dt
\end{equation}
%
%
%
%

Let's consider the simple case in which the Hamiltonian has just one critical
point. Taking the origin at the critical point and setting to zero the minimum
of the energy\footnote{This can always be done whenever, as we assume here,
the Hamiltonian is bounded from below. Indeed, shifting the Hamiltonian by a
constant: $H\rightarrow H+c$ will change the partition function by the
multiplicative constant $\exp(-\beta c)$. This implies that the only
thermodynamic quantity that will be affected by the shift will be the internal
energy: $U=:-\partial\ln\mathcal{Z}/\partial\beta$,\ which will be shifted by
$c$, leaving all the other thermodynamic functions unchanged $\cite{MO95}$,
and this is perfectly consistent. We will also assume the energy not to be
bounded from above.}we can evaluate the partition function as:
\begin{equation}
\mathcal{Z}=\underset{\underset{\varepsilon\rightarrow0}{E\rightarrow\infty}}
{\lim}\left\{ \int\limits_{S(E)}-\int\limits_{S(\varepsilon)}
\right\} \exp(-\beta H)\frac{dH\wedge dt}{h}
\end{equation}
where $S(E)$ (resp. $S(\varepsilon)$) is the volume in phase space bounded by
the constant energy surface $\Sigma(E)$ (resp. $\Sigma(\varepsilon)$). Using
then Stokes' theorem:
\begin{equation}
\mathcal{Z}=-\frac{1}{\beta h}\underset{\underset{\varepsilon\rightarrow
0}{E\rightarrow\infty}} {\lim}\left\{ \int\limits_{\Sigma(E)}\exp(-\beta
H)dt-\int\limits_{\Sigma(\varepsilon)}\exp(-\beta H)dt\right\}%
\end{equation}
and eventually:
\begin{equation}
\mathcal{Z=}-\frac{1}{\beta h}\underset{\underset{\varepsilon\rightarrow
0}{E\rightarrow\infty}}{\lim}\left\{ \exp(-\beta E)\tau(E)-\exp
(-\beta\varepsilon)\tau(\varepsilon)\right\}%
\end{equation}
and the desired result obtains immediately by taking the limits.
Generalizations to more than one critical point are easily obtained, and we
conclude that (''modulo'' an essentially irrelevant multiplicative factor
multiplicative factor, see footnote $6$ ) \textit{the partition function
depends only on the limiting value of the period \ ''at'' the critical points
of the Hamiltonian.} As the period is entirely determined by the dynamics, and
does not depend on the specific Hamiltonian description of the system, this
completes the proof in the more general case as well. \textit{\ }

We come now to the analogous problem in the context of Quantum Mechanics
$\cite{W50,MMZS97}$. In the spirit of the comments that we are making in this
Note, and to avoid mathematical complications arising from questions of
domains, convergence and the like, we will have in mind mainly a quantum
system on a finite-dimensional Hilbert space ( a spin, e.g.). This will allow
us nonetheless to highlight to some extent the r\^{o}le of the various
structures involved.

Let's consider then a complex separable Hilbert space $\mathcal{H}$ and and
Hamiltonian $\hat{H}$. The latter should be self-adjoint w.r.t. a given
Hermitian structure $<,>$ that, together with a linear structure\footnote{We
will consider here the linear structure as given once and for all, and
concentrate on the role of the Hamiltonian and of the Hermitian structure.},
qualifies $\mathcal{H}$ as a Hilbert space. From now on we will call \ $<,>$
the ''standard'' Hermitian structure on $\mathcal{H}$. We will assume the
spectrum of $\hat{H}$ to be discrete, and to correspond only to a finite
number of levels if necessary. Very loosely speaking (see however
Ref.$\cite{MMZS97}$ for a more accurate analysis involving also the
highlighting of the r\^{o}le of K\"{a}hler structures in Quantum Mechanics) on
going from Classical to Quantum Mechanics we replace the r\^{o}le of the pair
$(\Omega,H)$ with that of the pair $(<,>,\hat{H})$.

Adopting the Dirac notation, we will denote by $|\psi\rangle$ the state
\textit{vectors} (''\textit{kets}'' in Dirac's terminology) and by
$\langle\psi|$ their duals (the ''\textit{bras}'') w.r.t. the given Hermitian
structure. Notice that, as a \textit{''bra''} acts on vectors to give scalars,
the \textit{''bras}'' should be considered more properly as one-forms.
Consistently with this we will denote by $\psi_{k}$ the component of a given
ket $\psi$ \ on a given $O.N.$ basis \ $\{|k\rangle\}_{1}^{n}$ ($n=\dim
\mathcal{H}$, and hence: $\mathcal{H}\approx\mathbf{C}^{n}$ as a vector space)
and by $\psi^{k}$ the corresponding component of the associated bra, i.e.:%
\begin{equation}
\psi_{k}=\left\langle k|\psi\right\rangle ,\text{ \ }\psi^{k}=\left\langle
\psi|k\right\rangle
\end{equation}

\textit{Observables} will be represented (in the present context) by Hermitian
matrices: $A=A^{\dagger}$, and any one of them will be completely specified by
the equivalent assignment of the quadratic functional:
\begin{equation}
f_{A}(\psi)=:\langle\psi|A|\psi\rangle=\psi^{k}A_{k}\,^{l}\psi_{l}%
\end{equation}
where:  $A_{k}\,^{l}=\langle k|A|l\rangle$.

The Schr\"{o}dinger equation is written as:
\begin{equation}
i\hbar\frac{d}{dt}|\psi\rangle=\hat{H}|\psi\rangle
\end{equation}
or, setting $\hbar=1$ (or, equivalently, re-defining the Hamiltonian as
$\hat{H}/\hbar$) as:
\begin{equation}
i\frac{d\psi_{k}}{dt}=H_{k}\,^{l}\psi_{l}%
\end{equation}
All this is elementary and well known.

It is then easy to see $\cite{MMZS97}$ that the Schr\"{o}dinger equation can
be rewritten as the pair:
\begin{equation}
\frac{d\psi_{k}}{dt}=-i\frac{\partial f_{H}}{\partial\psi_{k}^{\ast}},\text{
\ }\frac{d\psi_{k}^{\ast}}{dt}=i\frac{\partial f_{H}}{\partial\psi_{k}}%
\end{equation}
Introducing now real coordinates $q_{k}$ and $p^{k}$ out of the complex
ones\footnote{And considering then now $\mathbf{C}^{n}\approx\mathbf{R}^{2n}$
as a real, $2n$-dimensional vector space}, the Schr\"{o}dinger equation can be
rewritten as a classical-like set of Hamiltonian equations $\cite{STR66}$,
i.e. as:
\begin{equation}
\frac{dq_{k}}{dt}=\frac{\partial f_{H}}{\partial p^{k}},\text{ \ }\frac
{dp^{k}}{dt}=-\frac{\partial f_{H}}{\partial q_{k}}%
\end{equation}
with $\ f_{H\text{ }}$ playing the role of the (classical-like) Hamiltonian.

The Hamiltonian vector field preserves then the Hermitian structure or,
equivalently, it is Hermitian w.r.t. the scalar product $<,>$.

Having cast the Schr\"{o}dinger equation into the Hamiltonian form $(30)$, it
is now clear that alternative quantum descriptions can be given by looking for
alternative Hermitian structures invariant under the Hamiltonian flow.

Indeed, a new scalar structure allows us to define a new quadratic function:
\begin{equation}
g_{A}(\psi)=:\langle\psi|A|\psi\rangle_{1}%
\end{equation}
and a new Poisson bracket (via the symplectic structure associated with the
imaginary part of the Hermitian structure). In this way we get the same
equations of motion in the form:
\begin{equation}
\frac{dq_{k}}{dt}=\frac{\partial g_{H}}{\partial p^{k}};\text{ \ }\frac
{dp^{k}}{dt}=-\frac{\partial g_{H}}{\partial q_{k}}%
\end{equation}
In deriving these equations we have assumed that the complex structure has
been unchanged.

If we consider now the trace of $H$, if it exists, or of any bounded operator
thereof like $\exp\{-\beta H\},$ we have (assuming discreteness):
\begin{equation}
\mathbb{I}=\sum_{n}|\psi_{n}\rangle_{1}\text{ }_{1}\langle\psi_{n}|=\sum
_{n}|\psi_{n}\rangle\langle\psi_{n}|
\end{equation}
and
\begin{equation}
Tr\exp\{-\beta H\}=\sum_{n}\langle\psi_{n}|\exp\{-\beta H\}|\psi_{n}%
\rangle_{1}=\sum_{n}\langle\psi_{n}|\exp\{-\beta H\}|\psi_{n}\rangle
\end{equation}
which is again what we wanted to prove. \bigskip

All this is recalled here in order to stress that, with some changes, what has
been said at the classical level can be carried over also to the
quantum-mechanical context. In other words, changing simultaneously the
Hamiltonian and the Hermitian structure in such a way that the description of
the dynamics w.r.t. the new pair $(\left\langle ,\right\rangle ,\widehat{H})$
be unchanged will lead also to the same statistical-mechanical description in
the appropriate ensembles, having in mind that the classical integration over
phase space w.r.t. the symplectic measure has to be replaced here by the
operation of \ taking traces, the latter depending on the chosen Hermitian structure.

We will consider now the analogous problem in the Heisenbeerg picture. Again
for the sake of simplicity, we will consider here only a simple example
related to the quantum one-dimensional harmonic oscillator ($QODHO$%
)\footnote{This will bring us of course beyond finite-dimensional Hilbert
spaces, but in a rather easily controlled context, though.}.

In terms of the creation and annihilation operators $a$ and $a^{\dagger}$,
with the standard commutation relations:%

\begin{equation}
\lbrack a,a^{\dagger}]=1
\end{equation}
one constructs a basis in the Fock space as:%

\begin{equation}
|n>_{1}=\frac{(a^{\dagger})^{n}}{\sqrt{n!}}|0\rangle
\end{equation}
with $|0\rangle$ the Fock vacuum and the standard scalar product, that we will
denote as $\langle.|.\rangle_{1}$:%

\begin{equation}
<n|m>_{1}=\delta_{nm}%
\end{equation}

The $QODHO$ is described by the Heisenberg equations of motion:
\begin{equation}
\frac{da}{dt}+ia=0
\end{equation}
together with the parent equation for $a^{\dagger}$, that are obtained in the
usual way from the Hamiltonian: $H=a^{\dagger}a+\frac{1}{2}$. \ Traces of
linear operators are defined then as:
\begin{equation}
Tr_{1}\hat{O}=\sum\limits_{n=0}^{\infty}\left\langle n|\hat{O}|n\right\rangle
_{1}%
\end{equation}
with $\hat{O}$ any trace-class operator, of course.

Now, we perform a ''nonlinear change of variables'' by defining $\cite{MMZS97}%
$ new operators as:
\begin{equation}
A=f(\widehat{n})a
\end{equation}
with $\ f(\widehat{n})$ a \ positive, monotonically increaasing and nowhere
vanishing function of the number operator $\widehat{n}=a^{\dagger}a$ (e.g.:
$f(.)=1+\tanh(.)$).

At this point, a little care is required when defining the adjoint of any
operator, as this notion depends in a crucial way on the \ hermitian structure
we are using on the Hilbert space of states. With the scalar product
$\left\langle .|.\right\rangle _{1}$, with which $a^{\dagger}$ is the adjoint
of $a$, the adjoint of $A$ is of course: $A^{\dagger}=a^{\dagger}f(\hat{n})$

It is pretty clear that \ $\widehat{n}$ \ being a constant of the motion the equations of motion for $A$ and
$A^{\dagger}$ will be the \ same as before. We can however reconstruct a different Fock space by \ assuming the
same vacuum and defining new states\footnote{Note that, with this definition:
$|n\rangle_{2}=\{\prod_{k=0}^{N-1}f(k)\}|n\rangle_{1}$} as:
\begin{equation}
|n>_{2}=\frac{(A^{\dagger})^{n}}{\sqrt{n!}}|0>
\end{equation}
with a \textbf{new} scalar product to be declared as:
\begin{equation}
<n|m>_{2}=\delta_{nm}%
\end{equation}

The nonlinearity of the transformation reflects itself in the fact that,
despite the fact that (see the previous footnote) \ $|n\rangle_{1}$and
\ $|n\rangle_{2}$ are proportional, the linear structure in the Fock space
labeled by $"1"$ does not carry over to the linear structure of space $"2"$.
This has to do with the fact that the proportionality factors between the
$|n\rangle_{1}$'s and the $|n\rangle_{2}$'s depend on $n$. In other words, if
we try to induce on space $"2"$ a linear structure modelled on that of
$\ $space $"1",$ the latter will not be compatible with the bilinearity of the
scalar product $\left\langle .|.\right\rangle _{2\text{ }}$ that we have just defined.

We may also wish to compute the commutator of the operators $A$ and
$A^{\dagger}$. This is easily done and we find:
\begin{equation}
\lbrack A,A^{\dagger}]=\Phi(\hat{n}+1)-\Phi(\hat{n})
\end{equation}
where: $\Phi(x)=xf^{2}(x)$.

Now, $A^{\dagger}$ is no more the adjoint of $A$ w.r.t. the new Hermitian
structure we have introduced. If \ we denote by $(.)_{2}^{\dagger}$ the
adjoint of any operator w.r.t. the second Hermitian structure, then we find:
\begin{equation}
(A^{\dagger})_{2}^{\dagger}=\frac{1}{f(\hat{n})}a
\end{equation}
which is quite different from $A$. The pair $\{(A^{\dagger})_{2}^{\dagger
},A^{\dagger}\}$ will yield a new (''nonlinear'') realization of the
Heisenberg algebra, and indeed it is immediate to see that:
\begin{equation}
\lbrack(A^{\dagger})_{2}^{\dagger},A^{\dagger}]=1
\end{equation}
Now, $(A^{\dagger})_{2}^{\dagger}$ and $A^{\dagger}$ will obey the same
equations of motion as $a$ and $a^{\dagger}$, that can be derived from the
previous commutation relations and from the Hamiltonian: $\widetilde
{H}=A^{\dagger}(A^{\dagger})_{2}^{\dagger}+1/2$ (which turns out actually to
coincide with the old one when written in terms of the original creation and
annihilation operators) and that will have therefore the same spectrum.
Defining then consistently the trace of any operator $\hat{O}$ as:
\begin{equation}
Tr_{2}\hat{O}=\sum\limits_{n=0}^{\infty}\left\langle n|\hat{O}|n\right\rangle
_{2}%
\end{equation}
will lead to the same partition function.

The examples discussed here seem therefore to point to the result that, both
in the classical as well as in the quantum case, the Hamiltonian description
is a sort of ''superstructure'' that is imposed on the thermodynamic
description of physical systems, and that what really matters are the
equations of motion of the systems themselves.

\end{document}